\documentclass[twocolumn,floatfix,amsmath,amssymb]{revtex4}
\usepackage{graphicx}
\usepackage{dcolumn}
\usepackage{color}

\newcommand{\be}{\begin{equation}}
\newcommand{\ee}{\end{equation}}

\begin{document}
\title{Effective diffusivity of passive scalars in rotating turbulence}
\author{P. Rodriguez Imazio$^1$ and P.D. Mininni$^{1,2}$}
\affiliation{$^1$ Departamento de F\'{\i}sica, Facultad de Ciencias Exactas y 
                  Naturales, Universidad de Buenos Aires and IFIBA, CONICET, 
                  Cuidad Universitaria, Buenos Aires 1428, Argentina \\
             $^2$ National Center for Atmospheric Research, P.O. Box 3000, 
                  Boulder, Colorado 80307, USA}
\date{\today}

\begin{abstract}
We use direct numerical simulations to compute turbulent 
transport coefficients for passive scalars in turbulent rotating
flows. Effective diffusion coefficients in the directions parallel and
perpendicular to the rotation axis are obtained by studying the diffusion 
of an imposed initial profile for the passive scalar, and calculated
by measuring the scalar average concentration and average spatial 
flux as a function of time. The Rossby and Schmidt numbers 
are varied to quantify their effect on the effective diffusion. It 
is find that rotation reduces scalar diffusivity in the perpendicular 
direction. The perpendicular diffusion can be estimated 
from mixing length arguments using the characteristic velocities and 
lengths perpendicular to the rotation axis. Deviations are observed 
for small Schmidt numbers, for which turbulent transport decreases 
and molecular diffusion becomes more significant.
\end{abstract}

\maketitle

\section{Introduction}

Turbulent transport in anisotropic flows plays an important role 
in a wide variety of astrophysical and geophysical processes 
(see, e.g., \cite{Schatzman,Charbonnel,Pipinyrudiger} for 
examples of transport in astrophysics, \cite{Roberts} for an 
example in the context of the geodynamo, \cite{Rotunno} for an 
example in atmospheric flows, and \cite{Osborn} for a study of 
vertical transport and diffusion in the ocean). In many of these 
examples, the resulting mixing is often studied and modeled by 
means of anisotropic transport coefficients 
\cite{Meneguzzi,Branden,Rotunno}.

While the transport of passive scalars in isotropic turbulent 
flows has received a lot of attention (see, e.g., 
\cite{Warhaft,Falkovich,Cencini}), less studies have 
considered the case of diffusion of the passive scalar in 
anisotropic flows. In the presence of anisotropy, the transport 
of the passive scalar is modified and turbulent transport 
coefficients should become anisotropic. Of particular relevance 
for the atmosphere and the oceans (as well as for mixing in 
stellar convective regions) are the cases or rotating and/or 
stratified turbulence. In the case of stratified flows, numerical 
simulations \cite{Meneguzzi} showed that the stratification 
acts to reduce the horizontal fluctuations of a scalar (i.e., it 
maintains the turbulent mixing in horizontal planes), while it 
suppresses the vertical mixing and transport of the passive 
scalar. These results were also confirmed analytically later in 
\cite{Pipin}.

Turbulent transport and diffusion of passive scalars in 
rotating flows has received less attention, although 
the anisotropy introduced by rotation is important at the 
largest scales of the atmosphere and the oceans, as well as 
in structures in the atmosphere that develop strong local 
rotation, as in tornadoes \cite{Bryan,Bryan2}. In the astrophysical 
context, the effects of rotation on turbulent diffusion are 
also important to understand the development of latitudinal 
entropy gradients in rotating stars \cite{Branden}. 
Furthermore, anisotropic passive scalar transport is believed 
to be associated with the partial depletion of lithium 
observed in the sun \cite{Pipinyrudiger}.

The development of anisotropy in rotating flows 
differs from that in stratified flows. 
While in the later case structures tend to be 
flat and layered, in rotating flows structures are elongated 
along the axis of rotation and the flow becomes quasi-two 
dimensional \cite{Liechtenstein}. In rotating 
flows, turbulent transport coefficients for passive scalars 
were obtained from numerical simulations using the test 
field model in \cite{Branden}. The results indicate that 
turbulent diffusion in the direction of rotation is reduced 
when the scales of the turbulence are comparable with 
those of the mean field. Recently, we considered scaling laws 
and intermittency of the passive scalar in rotating flows 
\cite{Imazio} and found that the distribution of the passive 
scalar becomes highly anisotropic, with scaling laws 
developing in the plane perpendicular to the rotation axis, 
and with scaling compatible with Kraichnan's model 
\cite{Kraichnan} for the passive scalar in two dimensions.

In a more general context, many authors have studied 
turbulent diffusion in anisotropic flows in terms of particle 
dispersion. These studies are related with the transport of 
passive scalars, as the average concentration of a passive 
scalar is related to single-particle dispersion in a fluid. 
Theoretical and numerical studies 
\cite{Vassilicos1,Cambon04} (see also 
\cite{Kaneda2,Kaneda1,Vassilicos2}, and simulations in 
\cite{Kimura1})  indicate one-particle dispersion is suppressed 
along the direction of stratification in stably stratified 
turbulence with and without rotation. Moreover, results in 
\cite{Cambon04} and \cite{Vassilicos2} show that the 
anisotropy introduced by stratification also reduces horizontal 
diffusion, although to a much lesser extent than vertical 
diffusion. These results are in good agreement with previous 
experimental results \cite{Britter}. In the case of rotating flows 
without stratification, the same studies indicate that 
horizontal diffusion is highly reduced while vertical diffusion 
remains nearly unchanged. More recently, Lagrangian 
velocity autocorrelations (related with turbulent diffusion) 
were obtained in experiments of forced rotating turbulence 
\cite{Castello}, and numerical simulations considered mixing in 
rotating flows \cite{Yeung}.

In this work, we study turbulent transport coefficients for the 
passive scalar in a turbulent rotating flow. Following a procedure 
similar to the one used in \cite{Meneguzzi} for stratified flows, 
effective coefficients are obtained by studying the diffusion of 
an initial distribution of the scalar quantity in a turbulent flow, 
and calculated by measuring the average concentration and 
average flux of the scalar. Unlike previous studies (see, e.g., 
\cite{Branden}), the coefficients obtained here are not scale 
dependent. Coefficients calculated in this fashion are useful 
when a computationally economical description of the problem 
is required, as is often the case, e.g., in weather and climate 
modeling \cite{Bryan}. As small-scale turbulence in 
weather simulations cannot be explicitly resolved (as an example, 
resolving small-scale turbulence in hurricanes would require 
grids with spacing smaller than 100 m \cite{Rotunno3}), and 
as subgrid models such as Large-Eddy Simulations require a 
fraction of the small-scale turbulence spectrum to be sufficiently 
resolved, a large fraction of numerical models in atmospheric 
sciences use simple turbulence parameterization schemes 
with effective diffusion and mixing coefficients \cite{Bryan3}. 
These numerical models are known to produce unphysical 
results if turbulence is not properly parameterized (see, e.g., 
\cite{Hausman,Persing}). Determination of anisotropic transport 
coefficients is thus important for such numerical and theoretical 
models.

In our study, the velocity field and the passive scalar 
concentration are obtained from direct numerical simulations 
(DNS) of the Navier-Stokes equations in a rotating frame,
together with the advection-diffusion equation for the scalar 
concentration. All DNS are performed using $512^3$ grid points 
in a regular periodic tridimensional grid. Three different rotation
rates are considered, chosen to study turbulence with moderate 
Rossby number. Diffusion coefficients are calculated for the 
vertical (i.e., parallel to the rotation axis) and horizontal 
directions. A more detailed study of turbulent diffusion is then 
made varying the Schmidt number, in order to observe Schmidt 
and P\`eclet number dependence. Finally, we present 
visualizations of the scalar which allow a more complete 
interpretation of the diffusive processes in the vertical and 
horizontal directions.

We find that rotation dramatically reduces scalar diffusivity in 
the horizontal direction (in comparison with the isotropic and 
homogeneous case). This is in good agreement with theoretical 
arguments in \cite{Vassilicos2} for single-particle dispersion, 
where it was found that vertical diffusion is twice larger than 
horizontal diffusion in the presence of pure rotation. Within 
error bars, our results are consistent with this finding. 
Moreover, we find that horizontal diffusion can be reasonably 
estimated using mixing length arguments as $U_\perp L_\perp$ 
(where $U_\perp$ and $L_\perp$ are respectively the 
characteristic velocity and lengthscale in the direction 
perpendicular to the rotation axis). For small Schmidt and 
P\`eclet numbers, the turbulent diffusion decreases as 
molecular diffusion becomes more important, while for large 
enough P\`eclet numbers the effective diffusion becomes 
independent of the P\`eclet number.

\section{Numerical simulations}

\subsection{Equations and parameters}

Data analyzed in the following section stems form DNS of the 
incompressible Navier-Stokes equations in a rotating frame 
for the velocity ${\bf u}$, and of the advection-diffusion 
equation for the passive scalar $\theta$, given by
\begin{equation}
\partial_t {\bf u} + {\bf u}\cdot \nabla {\bf u} = -2{\bf \Omega} 
    \times {\bf u} - \nabla p + \nu \nabla^2 {\bf u} +{\bf f},
\label{eq:NS}
\end{equation}
\begin{equation}
\nabla \cdot {\bf u} =0, 
\label{eq:incomp}
\end{equation}
\begin{equation}
\partial_t {\theta} + {\bf u}\cdot \nabla {\theta} =  
    \kappa \nabla^2 {\theta}.
\label{eq:theta}
\end{equation}
Here $p$ is the pressure divided by the mass density (chosen 
uniform and constant in all simulations), $\nu$ is the kinematic 
viscosity, and $\kappa$ is the scalar diffusivity. Also, ${\bf f}$ 
is an external force that drives the turbulence, and 
${\bf \Omega} = \Omega \hat{z}$ where $\Omega$ is the 
rotation angular velocity. The mechanical forcing ${\bf f}$ is a 
superposition of Fourier modes with random phases, 
delta-correlated in time, injected in a narrow band of 
wavenumbers $k \in [1,2]$ (therefore we will consider the forcing 
wavenumber as $k_F \approx 1$).

Equations (\ref{eq:NS}), (\ref{eq:incomp}) and (\ref{eq:theta}) 
are solved in a three dimensional domain of size $2\pi$ with 
periodic boundary conditions using a parallel pseudospectral 
code \cite{Gomez05a,Gomez05b}. The pressure is obtained by 
taking the divergence of Eq.~(\ref{eq:NS}), using the 
incompressibility condition (\ref{eq:incomp}), and solving the 
resulting Poisson equation. The equations are evolved in time 
using a second order Runge-Kutta method. The code uses the 
$2/3$-rule for dealiasing, and as a result the maximum 
wavenumber is $k_{max} = N/3$, where $N$ is the number of 
grid points in each direction (with $N=512$ in all the runs). All 
simulations presented are well resolved, in the sense that the 
dissipation wavenumbers $k_\nu$ and $k_\kappa$ are smaller than 
the maximum wavenumber $k_{max}$ at all times.

The dimensionless numbers used to characterize the runs are 
the Reynolds, P\`eclet, and Rossby numbers, defined respectively 
as
\begin{equation}
R_e=\frac{UL}{\nu},
\end{equation}
\begin{equation}
P_e= S_c R_e = \frac{\nu}{\kappa} R_e,
\end{equation}
\begin{equation}
R_o = \frac{U}{2L\Omega},
\end{equation}
where $U$ is the r.m.s.~velocity in the turbulent steady state, and 
$L$ is the forcing scale of the flow defined as $L=2\pi/k_F$. $S_c$ 
is the Schmidt number, defined as $S_{c}= \nu / \kappa$. In all 
simulations, $U \approx 1$, and the viscosity is 
$\nu=6 \times 10^{-4}$. The passive scalar diffusivity $\kappa$ is 
set equal to the viscosity for the main set of runs, although similar 
simulations but changing the value of $S_{c}$ (and therefore of 
$P_e$) were also performed.

The detailed procedure followed in the numerical simulations is as 
follows. We first conducted a simulation solving Eqs. (\ref{eq:NS}) 
and (\ref{eq:incomp}) (Navier-Stokes without passive scalar and 
without rotation), starting from the fluid at rest (${\bf u} = 0$),
and applying the external forcing ${\bf f}$ until reaching a turbulent 
steady state. This run was continued for approximately $13$ 
turnover times. The final state of this run was used as initial 
condition for the velocity field in multiple runs in which the
external forcing ${\bf f}$ was kept the same but a passive scalar was
injected. These runs can be grouped (namely in sets $A$, $B$, $C$, 
and $E$), with each set corresponding to several runs with the 
same Rossby number (see table \ref{table:runs}). 

\begin{table}
\caption{\label{table:runs}Parameters used in each set of runs. 
              $\Omega$ is the rotation rate, $R_o$ is the Rossby
              number, $R_e$ is the Reynolds number, and $U$ is 
              the r.m.s. velocity in the turbulent steady state.}
\begin{ruledtabular}
\begin{tabular}{ccccc}
Set &   $\Omega$ & $R_o$ &     $R_e$  & $U$   \\
\hline
$A$ &        $0$      &   $\infty$ &   $1050$  &  1\\
$B$ &        $2$      &   $0.04$   &  $1050$  &   1\\
$C$ &        $4$      &   $0.02$ &  $1050$   &  1\\
$E$ &       $8$      &   $0.01$   &   $1050$    & 1
\end{tabular}
\end{ruledtabular}
\end{table}

\begin{table}
\caption{\label{table:runs2}Parameters used in one of the subsets 
               of runs. In the name of each run, the subindex $x$ or
               $z$ indicates the dependence with the 
               coordinate of the initial passive scalar 
               Gaussian profile, and the subindex with
               the number is $1/S_c$. For each run, $\nu$ is the
               kinematic viscosity, $\kappa$ is the molecular
               diffusivity, $P_e$ is the P\`eclet number, and $S_c$ is
               the Schmidt number.}
\begin{ruledtabular}
\begin{tabular}{cccccc}
Run &   $\nu$ & $\kappa$ &     $P_e$      & $S_c$     \\
\hline
$A_{x1/2}$&        $6 \times 10^{-4}$      &  $3 \times 10^{-4}$ &   $2100$  &  $2$ \\
$A_{x1}$  &        $6 \times 10^{-4}$      &  $6 \times 10^{-4}$   &  $1050$   &  $1$\\
$A_{x2}$  &       $6 \times 10^{-4}$      &  $1.2 \times 10^{-3}$ &  $520$   &   $0.5$\\
$A_{x4}$  &       $6 \times 10^{-4}$     &   $2.4 \times 10^{-3}$   &  $260$ &  $0.25$\\
$A_{x8}$  &       $6 \times 10^{-4}$     &   $4.8 \times 10^{-3}$   &   $130$  & $0.12$\\
$A_{x16}$&       $6 \times 10^{-4}$     &   $9.6 \times 10^{-3}$   &   $66$  & $0.06$\\
$A_{x32}$&       $6 \times 10^{-4}$     &   $1.92\times 10^{-2}$   &   $33$ &  $0.03$
\end{tabular}
\end{ruledtabular}
\end{table}

Each run in each set corresponds to a simulation in which an 
initial Gaussian profile for the passive scalar was injected as
follows:
\begin{equation}
\theta(t=0,x_i) = \theta_0 e^{- (x_i-\mu)^{2}/\sigma^{2}}
\end{equation}
where $i=1$ or 3 (i.e., the initial profile is a function of $x_1=x$ or
$x_3=z$),  $\mu =\pi$ (the profile is centered in the middle of the
box, with the box of length $2\pi$), and $\sigma=1$. This allows 
us to study the diffusion of the initial profile in the direction
parallel to rotation ($z$, or vertical) and in the direction
perpendicular ($x$, or horizontal). For a few runs it was explicitly 
verified that the diffusion in the $x$ and $y$ directions was the same 
(as rotating turbulence tends to be axisymmetric). The runs in each set 
are labeled with a subindex indicating the dependence of the initial 
profile (e.g., runs in group $A$ are labeled $A_{x}$ or $A_{z}$ depending 
on the initial Gaussian profile used). 

Finally, to measure the effect of varying the Schmidt number on the 
effective diffusivity, all simulations in each group (with subscript
$x$ or $z$) were repeated with different values of $\kappa$. In 
practice, we performed simulations increasing the molecular
diffusivity from $\kappa = \nu$ (all the simulations described above) to 
$\kappa = 32 \nu$ (increasing $1/S_c$ by factors of two for each
run). Simulations with $\kappa = \nu/2$ were also performed. This 
results in twelve more simulations in each group, with $S_c=2$ to 
$S_{c}=1/32$. To differentiate each run in each set, a 
subindex equal to $1/S_c$ is added after the subindex $x$ or $z$. 
The resulting list of all runs in set A is shown in table \ref{table:runs2}, 
indicating the value of $\nu$, $\kappa$, and the corresponding $P_{e}$ 
and $S_{c}$ numbers. The runs in the other sets are labeled following 
the same rules, and have the same parameters as the ones shown in 
Table \ref{table:runs2}. As an example, runs $A_{x4}$  and $C_{x4}$ 
have the same values of $\nu$, $\kappa$, $P_{e}$, and $S_{c}=1/4$, 
with an initial Gaussian profile for the passive scalar in $x$. The
two runs differ just in the Rossby number.

Since for each set of values of $R_0$ we considered $7$ different 
Schmidt numbers, and since for each of these set of parameters we 
performed two runs (one for the initial profile of the passive scalar
in the $x$ direction, and one for the profile in the $z$ direction), the 
total number of numerical simulations analyzed below is $56$.

\subsection{Turbulent diffusion coefficients}

To characterize the turbulent diffusion of the passive scalar, we
consider quantities averaged over the two directions perpendicular 
to the initial dependence of the Gaussian profile. In particular, we 
consider the average passive scalar concentration $\overline{\theta}$ 
and the spatial flux $\overline{\theta u_{i}}$, where $i=1$ or 3 
depending on the initial dependence of the Gaussian profile, and the 
averages are done over the two remaining Cartesian coordinates. 
Note the spatial flux $\overline{\theta u_{i}}$ represents the amount 
of passive scalar transported in the $i$-direction per unit of time by 
the fluctuating (or turbulent) velocity. Since there is no mean flow
in the simulations (we use delta-correlated in time random-forcing), 
$u_{i}$ is the fluctuating velocity. Then, the pointwise effective 
turbulent diffusion coefficient at $x_i$ and at a given time $t$ is 
\cite{Meneguzzi}
\begin{equation}
{\cal D}_{i}(x_i,t)=
    \frac{\overline{\theta u_{i}}}{\partial _{x_i} \overline{\theta}}.
\label{eq:DT}
\end{equation}
This coefficient corresponds to how much passive scalar is 
transported by the fluctuating velocity, per unit of variation of
$\overline{\theta}$ with respect to $x_i$. As already mentioned, $i=1$
will stand for horizontal diffusion, while $i=3$ will stand for vertical
diffusion.

In nature, turbulence often results from an instability of a 
large-scale flow (e.g., convection in stars, in the Earth's core and 
in the atmosphere, or the general circulation at the largest scales 
in the atmosphere and the oceans). The random forcing described 
above is intended to mimic the behavior of rotating turbulence at 
scales smaller than the large-scale flow; together with the rotation 
it produces an anisotropic flow with column-like structures, but it 
doesn't produce a coherent mean flow. When such a mean flow is 
present, Eq.~(\ref{eq:DT}) remains valid but the fluctuating velocity 
has to be defined from the total velocity field {\bf v} after
subtracting the mean flow
\begin{equation}
{\bf u} = {\bf v} - \left<{\bf v}\right> ,
\end{equation}
where the mean flow $\left<{\bf v}\right>$ can be defined, e.g., from 
a time average, in such a way the mean value of the fluctuating 
velocity, $\left<{\bf u}\right>$, is zero.

From Eq.~(\ref{eq:DT}), we can define averaged diffusion 
coefficients as follows. We can first average over the coordinate 
$x_i$ to obtain a time dependent turbulent diffusion,
\begin{equation}
{\cal D}_i(t) = \frac{1}{2 \pi} \int_0^{2\pi} {\cal D}_i(x_i,t) \mbox{d}x_i ,
\end{equation}
and we can further average over time, to obtain the mean turbulent 
diffusion
\begin{equation}
{\cal D}_i = \frac{1}{T} \int_{t_0}^{t_0+T} {\cal D}_i(t) \mbox{d}t .
\label{eq:taverage}
\end{equation}
Here, $t_0$ and $T$ are characteristic times of the flow. In practice, 
in our simulations the pointwise turbulent diffusion coefficients 
${\cal D}_i(x_i,t)$ first grow in time, as the initial Gaussian profile is 
mixed by the turbulence, reach a maximum, and then decrease as the 
scalar becomes diluted (which happens in most runs after at most three 
turnover times). This is reminiscent of the behavior of the energy 
dissipation in freely decaying turbulence, which reaches a maximum 
and then decreases. Following a standard practice when studying 
decaying flows, in the runs we chose $t_0$ as the time of saturation 
in the growth of ${\cal D}_i(t)$, and $T\approx 1$, proportional to the 
turnover time.

\begin{figure}
\centerline{\includegraphics[width=8.3cm]{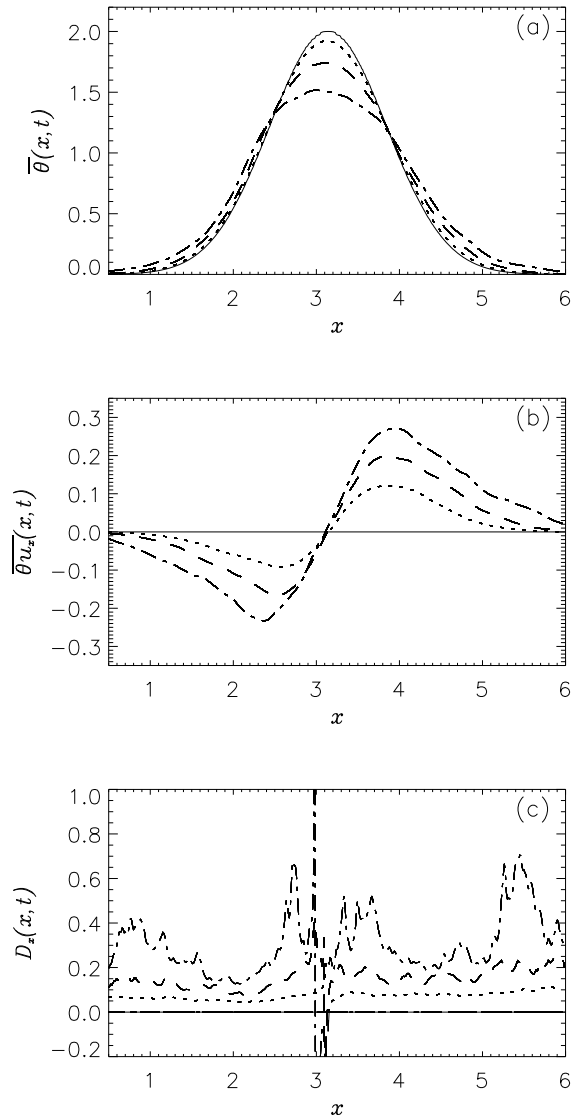}}
\caption{(a) Average horizontal concentration $\overline{\theta}(x,t)$ 
    in run $A_{x1}$, at times $t=0, 0.5, 1$ and $1.5$,
    denoted by solid, dotted, dash, and dash-dotted
    lines respectively. (b) Horizontal flux at the same times.
    (c) Pointwise horizontal turbulent diffusion at the same times.}
\label{fig:fig1}
\end{figure}

\section{Numerical results}

\subsection{Isotropic turbulence}

In the absence of rotation (runs in set A), we expect the diffusion
coefficients to be isotropic (i.e., the horizontal and vertical
diffusions to be the same within error bars). Figure \ref{fig:fig1} 
shows the horizontal average profile of the passive scalar 
$\overline{\theta}(x,t)$, the horizontal flux $\overline{\theta u_x}(x,t)$,
and the pointwise horizontal diffusion ${\cal D}_{x}(x,t)$ at five
different times in run $A_{x1}$. As time evolves, the mean profile 
$\overline{\theta}(x,t)$ flattens and widens. The flux is roughly 
antisymmetric, and is positive for $x>\pi$ and negative for $x<\pi$. 
This can be expected as there is an excess of $\theta$ at $x=\pi$ at 
$t=0$ which turbulent mixing should diffuse, by transporting this 
excess towards $x=0$ and $x=2\pi$. Horizontal diffusion increases 
to its saturation value around $t_0\approx 1.5$; after this time it 
fluctuates around a mean value. Large fluctuations in ${\cal D}_{x}(x,t)$ 
near the center of the box are due to the fact that, by definition, 
turbulent diffusion diverges at that point. 

Passive scalar concentration, vertical flux and vertical turbulent 
diffusion were also calculated for run $A_{z1}$ (i.e., the same run
but with an initial Gaussian profile in $z$). As expected, the same 
results as in run $A_{x1}$ were obtained.

\begin{figure}
\centerline{\includegraphics[width=8.3cm]{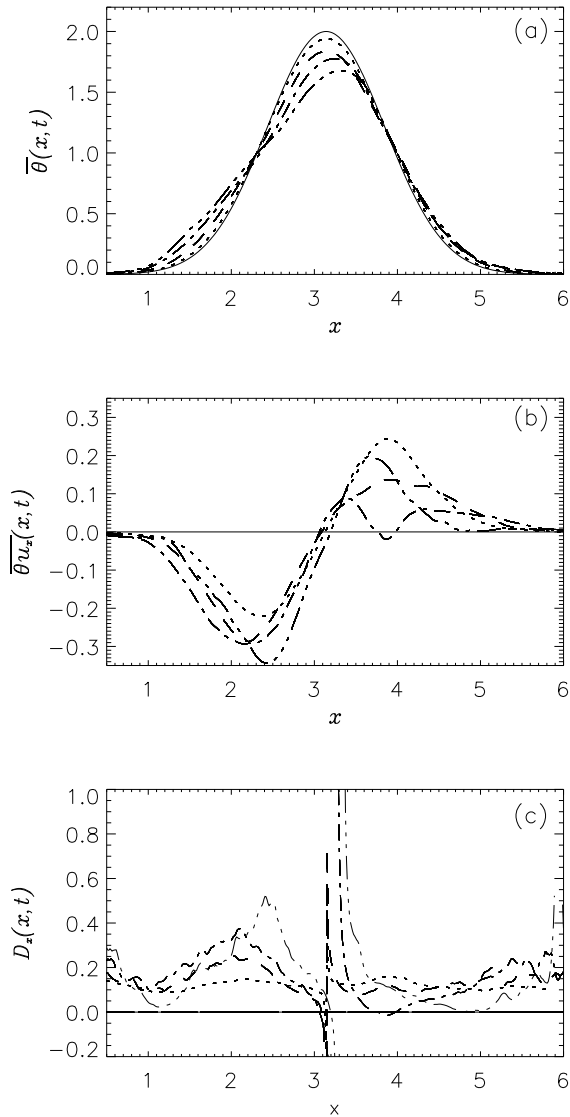}}
\caption{(a) Averaged vertical concentration $\overline{\theta}(x,t)$ 
    in run $C_{x1}$, at times $t =0, 0.25, 0.5, 0.75,$ and $1$, 
    denoted by solid, dotted, dash, dash-dotted and
    dash-triple-dotted lines respectively.
    (b) Horizontal flux at the same times. 
    (c) Pointwise horizontal turbulent diffusion at the same times.}
\label{fig:fig2}
\end{figure}

\subsection{Effect of rotation}

\subsubsection{Horizontal transport}

Figure \ref{fig:fig2} shows the mean horizontal concentration 
$\overline{\theta}(x,t)$, the horizontal flux $\overline{\theta u_x}(x,t)$,
and the pointwise horizontal diffusion ${\cal D}_{x}(x,t)$ for different 
times in run $C_{x1}$ ($\Omega=4$). In this case diffusion grows faster 
at earlier times, but then saturates at a lower value. Also note that the 
average profile and the average flux become asymmetric: there is an 
excess of $\overline{\theta}(x,t)$ for $x<\pi$, and the flux (in absolute 
value) is larger for $x<\pi$ than for $x>\pi$. This asymmetry is 
associated with the Coriolis force and has been already observed in
\cite{Branden}. As will be shown later, it results from a rotation of
the initial passive scalar profile.

\begin{figure}
\centering
\includegraphics[width=8.3cm]{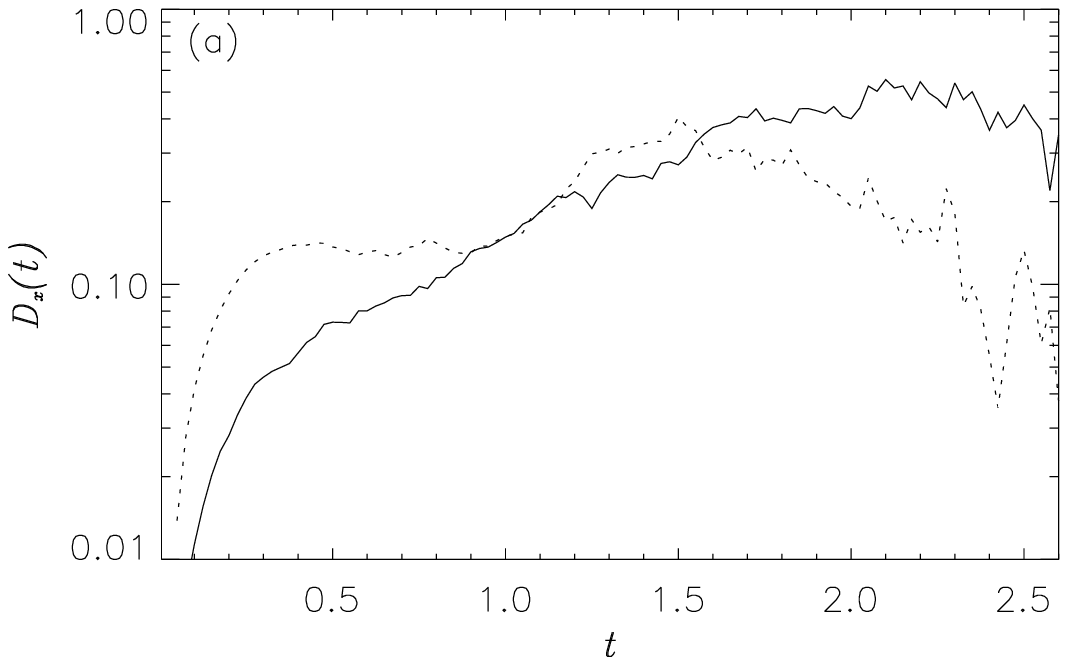}\\
\includegraphics[width=8.3cm]{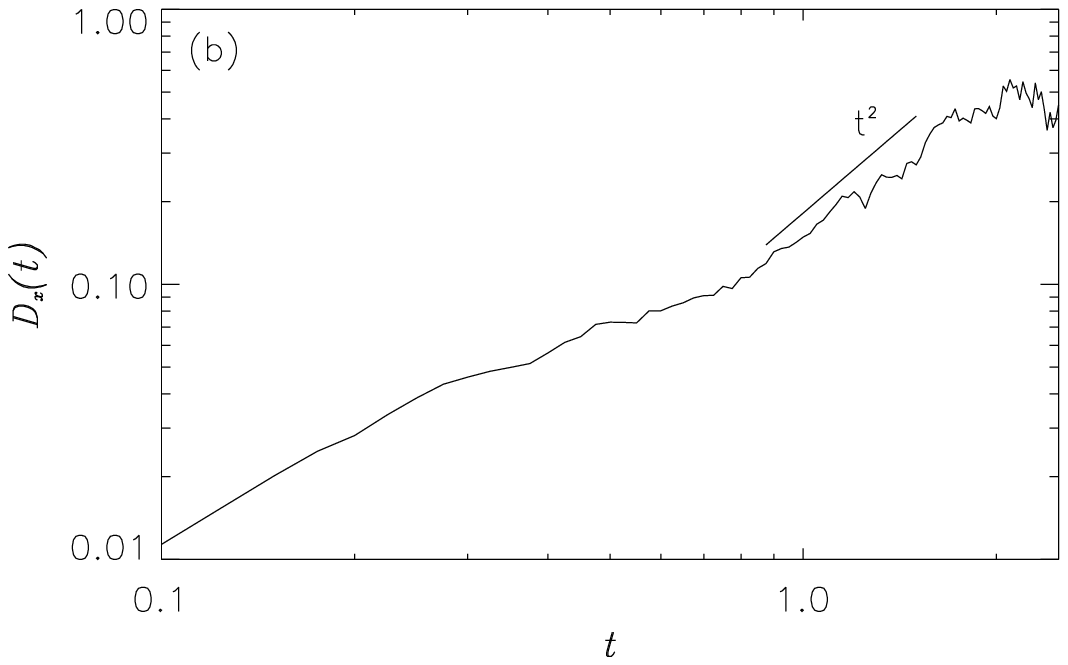}\\
\includegraphics[width=8.3cm]{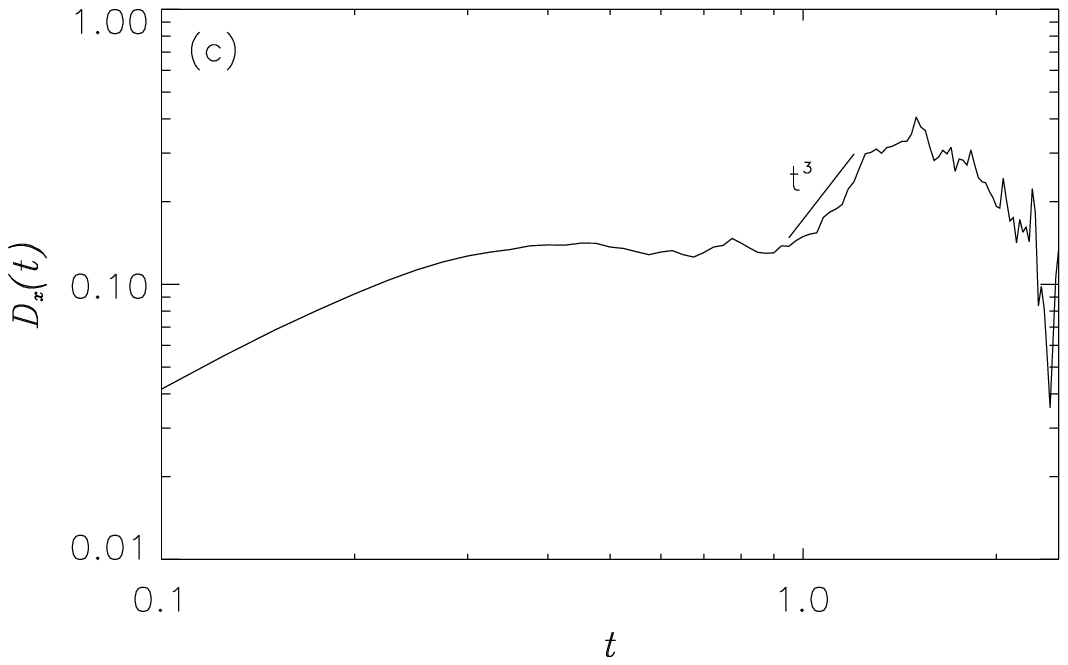}
\caption{(a) Horizontal turbulent diffusion as a function of time for 
    runs $A_{x1}$ (solid) and $C_{x1}$ (dotted).
    (b) Detail in log-log coordinates of the horizontal turbulent 
    diffusion for run $A_{x1}$.
    (c) Same for run $C_{x1}$.}
\label{fig:fig3}
\end{figure}

From ${\cal D}_{x}(x,t)$ calculated for each time, we obtain the 
turbulent diffusion coefficient ${\cal D}_{x}(t)$ by computing its mean 
over all values of $x$. Figure \ref{fig:fig3} shows the resulting 
horizontal turbulent diffusion as a function of time for runs $A_{x1}$ 
and $C_{x1}$. In both runs ${\cal D}_{x}(t)$ grows from an initially small
value to a saturation value around $t_0\approx1.5$ in run 
$A_{x1}$, and $t_0 \approx 1.2$ in run $C_{x1}$. These two values 
are typical of all runs without and with rotation respectively, and 
were used to compute the time average of ${\cal D}_{x}(t)$ using 
Eq.~(\ref{eq:taverage}). 
Note that $t_0$ is smaller in run $C_{x1}$, as the runs with rotation 
show a faster growth of ${\cal D}_{x}(t)$ at early times, but saturate 
faster at a lower value. Later, ${\cal D}_{x}(t)$ decreases as the 
passive scalar is diffused and homogeneity is finally recovered. As can 
be seen in Fig.~\ref{fig:fig3}(a), this decrease is more evident in run 
$C_{x1}$.

As already mentioned, although at early times ${\cal D}_{x}(t)$ grows 
faster in the run with rotation (run $C_{x1}$), ${\cal D}_{x}(t)$ saturates 
at a lower value than in run $A_{x1}$. This indicates horizontal diffusion 
decreases in the presence of rotation, in comparison with the 
isotropic and homogeneous case. This is in good agreement with 
predicted reductions of the horizontal passive scalar transport in 
the presence of rotation \cite{Kimura1,Vassilicos2}, and also with 
previous measurements of turbulent transport coefficients 
in numerical simulations using a different method (the test field
model, see \cite{Branden}).

\begin{figure}
\centerline{\includegraphics[width=8.3cm]{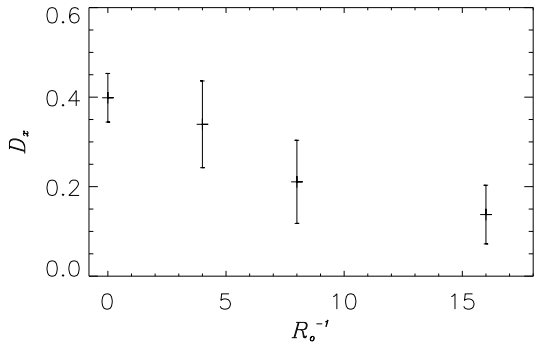}}
\caption{Mean horizontal turbulent diffusion as a function of the 
    inverse Rossby number, obtained from simulations $A_{x1}$, 
    $B_{x1}$, $C_{x1}$, and $E_{x1}$ (all runs with $S_{c}=1$). Error bars
    correspond to the mean standard deviation.}
\label{fig:fig4}
\end{figure}

It is interesting that at intermediate times the growth of ${\cal D}_{x}(t)$
can be roughly explained by a simple phenomenological model. 
If we think that turbulent eddies dominate the transport over 
molecular diffusion, and we think of the turbulent flow as a
superposition of eddies at different scales $\ell$, then we can
imagine that the smallest eddies (much smaller than the integral scale
eddies), with faster turnover times $\tau_\ell$, are the first to
start mixing the passive scalar. As time evolves, larger and larger 
eddies come into play, as the eddies at larger scales are able to complete 
a turnover. For eddies in the inertial range, the eddy turnover time
can be estimated as 
\begin{equation}
\tau_\ell \sim \ell/u_\ell
\label{eq:tau1}
\end{equation}
In isotropic and homogeneous turbulence, the inertial range scaling 
for the velocity field (in the absence of intermittency corrections)
can be written as $u_\ell \sim \ell^ {1/3}$. Then, the eddy turnover
time is
\begin{equation}
\tau_\ell \sim \ell^{2/3}.
\label{eq:tau2}
\end{equation}
From mixing length arguments, the turbulent diffusion at late times 
is ${\cal D}_{x} \sim LU$. At early times, if only the eddies with turnover
time smaller than the actual time contribute to the mixing, then from
Eqs.~(\ref{eq:tau1}) and (\ref{eq:tau2}) we get
\begin{equation}
{\cal D}_{x} \sim \tau_\ell^{2}.
\label{eq:DT1}
\end{equation}

\begin{figure}
\centerline{\includegraphics[width=8.3cm]{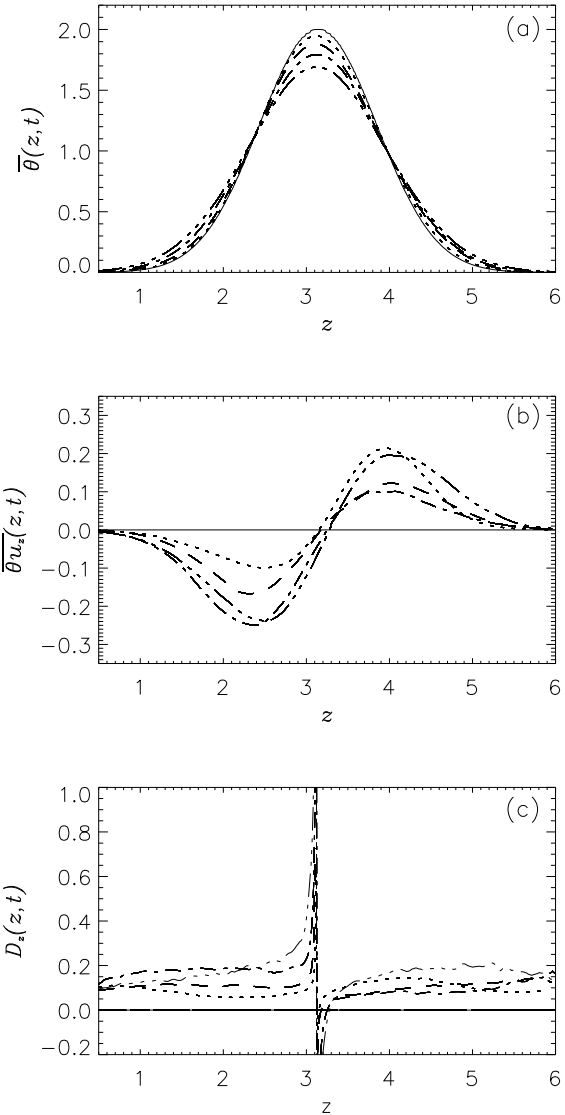}}
\caption{(a) Averaged passive scalar profile $\overline{\theta}(z,t)$ 
    in run $C_{z1}$, at times $t = 0, 0.25, 0.5, 0.75,$ and $1$,
    denoted by solid, dotted, dash, dash-dotted and
    dashed-triple-dotted lines respectively.
    (b) Mean vertical flux as a function of $z$ at different times in 
    the same run. (c) Same for the pointwise vertical turbulent diffusion.}
\label{fig:fig5}
\end{figure}

In the rotating case the inertial range scaling for the velocity field is
modified. Rotation sets a preferential direction for energy transfer, 
resulting in a quasi-bidimensionalization of the flow. Assuming 
$E(k_{\perp}) \sim k_{\perp}^{-2}$ in the inertial range (see, e.g., 
\cite{Cambon,Mininni09}), it follows that 
$u_{\ell_{\perp}} \sim \ell_{\perp}^{1/2}$. If at late times we can
assume the horizontal diffusion ${\cal D}_{x} \sim L_\perp U_\perp$ (with 
$L_\perp$ and $U_\perp$ respectively the characteristic length and 
velocity in the perpendicular direction), and using the same 
arguments as in the isotropic and homogeneous case, then
\begin{equation}
{\cal D}_{x} \sim \tau_\ell^{3} .
\label{eq:DT2}
\end{equation}
Assuming that $t\sim \tau_\ell$ (i.e., that at a given time only the 
eddies that were able to do a full turn contribute to the mixing), one 
could expect to observe the growth given by Eqs.~(\ref{eq:DT1}) and 
(\ref{eq:DT2}) before saturation in ${\cal D}_{x}(t)$. These scaling 
laws are indicated as a reference in Figs.~\ref{fig:fig3}(b) and
(c). Although there is an agreement with the prediction, the ranges 
are shorter than a decade and we cannot conclude that the scalings 
appear in the simulations. Also, the behavior of run $A_{x1}$ at 
early times can be also compatible with exponential growth of 
${\cal D}_{x}(t)$. Given the large fluctuations in ${\cal D}_{x}(t)$
in all runs, more simulations with the same parameters (but with 
different initial conditions to perform an ensamble average) would be 
needed to study this early time behavior in more detail.

The reduction of the saturation value of the horizontal turbulent 
diffusion observed in Figs.~\ref{fig:fig1}, \ref{fig:fig2}, and
\ref{fig:fig3} can be further confirmed by studying the mean 
temporal value of ${\cal D}_{x}(t)$ (namely ${\cal D}_{x}$) for all
runs with $S_c=1$. Figure \ref{fig:fig4} shows the mean temporal 
value of the horizontal diffusion as a function of the Rossby number 
for runs $A_{x1}$, $B_{x1}$, $C_{x1}$, and $E_{x1}$ (i.e., simulations 
with $\Omega = 0, 2, 4$ and $8$, and with initial Gaussian profile 
of the scalar in $x$). The error bars correspond to the mean standard 
deviation. It is clear from the data that ${\cal D}_{x}$ decreases 
monotonically as $1/R_o$ increases.

\subsubsection{Vertical diffusion}

\begin{figure}
\centerline{\includegraphics[width=8.3cm]{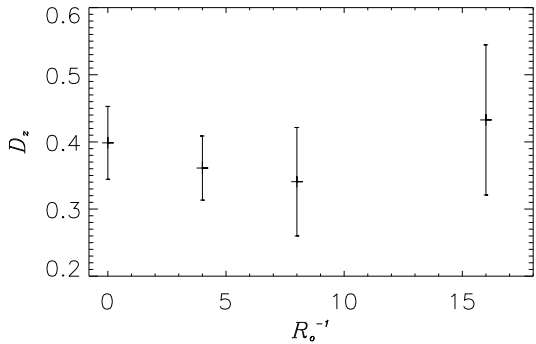}}
\caption{Mean vertical turbulent diffusion as a function of the 
    inverse Rossby number, obtained from simulations 
    $A_{z1}$, $B_{z1}$, $C_{z1}$, and $E_{z1}$ (all runs with
    $S_c=1$). Error bars correspond to the mean standard deviation.}
\label{fig:fig6}
\end{figure}

\begin{figure}
\centerline{\includegraphics[width=8.3cm]{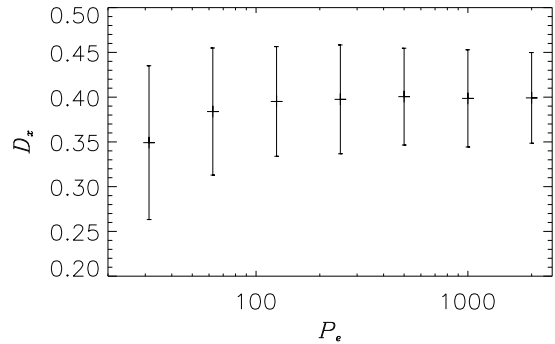}}
\caption{Mean turbulent diffusion ${\cal D}_{x}$ as a function of the 
    P\`eclet number for  runs  $A_{x1}$ to $A_{x32}$ ($\Omega=0$). 
    Note that as the flow in these runs is approximately isotropic, 
    ${\cal D}_{x}\approx {\cal D}_{z}$.}
\label{fig:fig7}
\end{figure}

Figure \ref{fig:fig5} shows the mean vertical passive scalar
concentration $\overline{\theta}(z,t)$, the mean vertical flux 
$\overline{\theta v_{z}}(z,t)$, and the pointwise vertical diffusion 
${\cal D}_{z}(z,t)$ at different times in run $C_{z1}$. Note the profiles 
here are more similar to those obtained in the isotropic and 
homogeneous case: $\overline{\theta}(z,t)$ and 
$\overline{\theta v_{z}}(z,t)$ are respectively symmetric and 
antisymmetric with respect to $z=\pi$. 

As in the case of horizontal diffusion, we can compute the mean 
vertical diffusion coefficient in runs $A_{z1}$, $B_{z1}$, $C_{z1}$,
and $E_{z1}$. This results in a dependence of ${\cal D}_{z}$ with the 
Rossby number, and is shown in Fig. ~\ref{fig:fig6}. While mean 
horizontal diffusion is strongly dependent on the Rossby number, 
the mean vertical diffusion seems to be (within error bars) 
independent of its value.

\subsection{Effect of $S_c$ and $P_e$ numbers}

\begin{figure}
\centerline{\includegraphics[width=8.3cm]{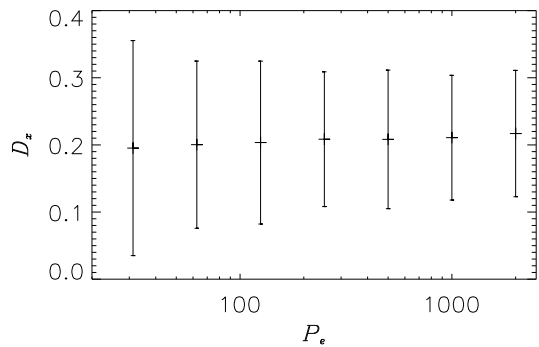}}
\caption{Mean horizontal turbulent diffusion as a function
    of the P\`eclet number for  runs $C_{x1}$ to $C_{x32}$ 
    ($\Omega = 4$).}
\label{fig:fig8}
\end{figure}

\begin{figure}
\centerline{\includegraphics[width=8.3cm]{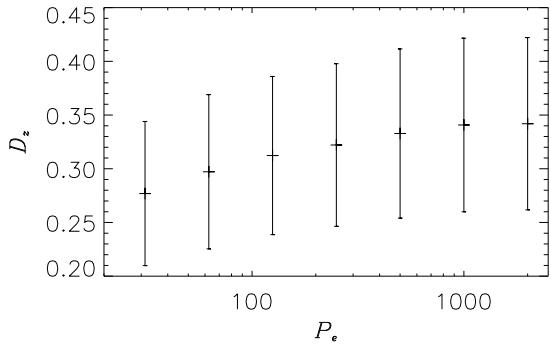}}
\caption{Mean vertical turbulent diffusion as a function of the 
    P\`eclet number for runs $C_{z1}$ to $C_{z32}$ ($\Omega = 4$).}
\label{fig:fig9}
\end{figure}

A large number of runs were performed to study the effect of 
the Schmidt number in turbulent diffusion. As the molecular 
diffusivity $\kappa$ is increased, we can expect turbulent diffusion 
to decrease until molecular diffusion dominates. Figure \ref{fig:fig7} 
shows the mean turbulent diffusion ${\cal D}_{x}$ as a 
function of $P_{e}$ in runs with $\Omega=0$ (i.e., estimating 
${\cal D}_{x}$ from all the runs in set $A$). As the flow is 
approximately isotropic, ${\cal D}_{x} \approx {\cal D}_{z}$.

Effects associated with the P\`eclet and Schmidt numbers can be 
observed for small $P_e$, as ${\cal D}_{x}$ starts to decrease for 
$P_e \lesssim 100$. However, for a wide range of values of the 
P\`eclet number ${\cal D}_{x}$ remains approximately constant. This 
is in good agreement with theoretical expectations that for small 
enough $\nu$ and $\kappa$, the effective turbulent Schmidt 
number should be of order one (see \cite{Ponty} for similar arguments 
in the case of the turbulent magnetic Prandtl number).

\begin{figure}
\includegraphics[width=8.3cm]{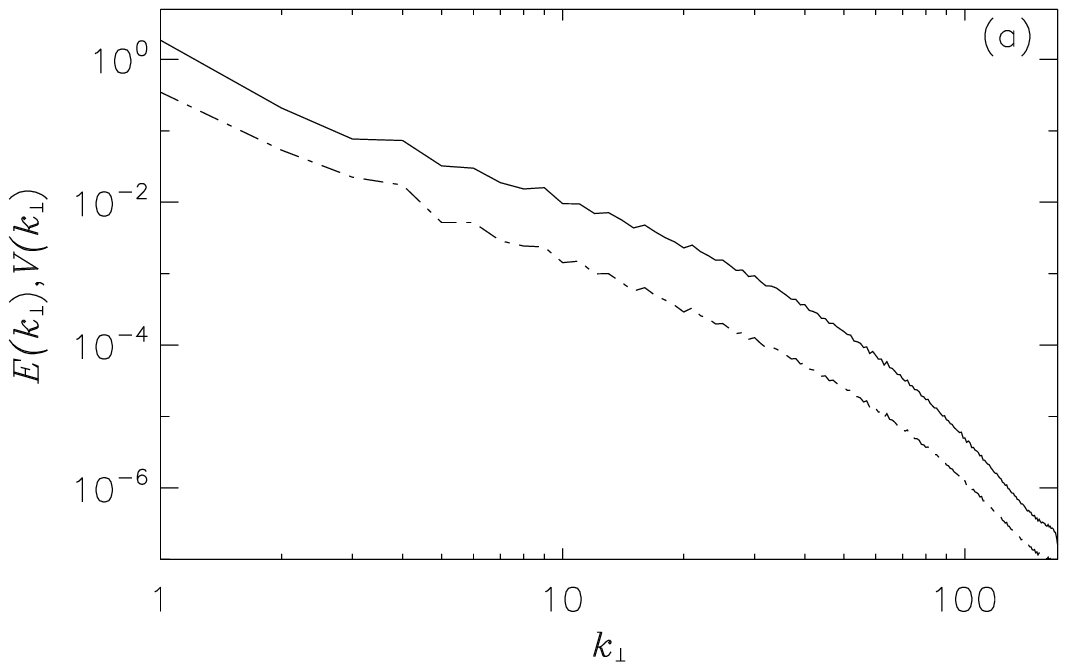}
\includegraphics[width=8.3cm]{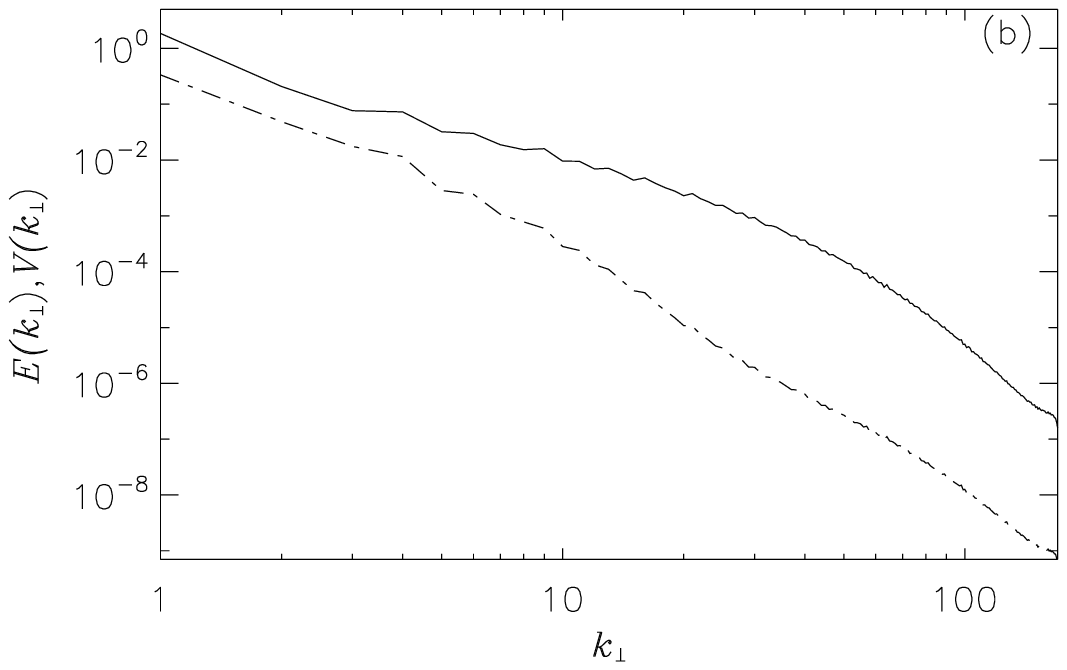}
\caption{(a) Horizontal energy spectrum (solid) and passive scalar 
    variance spectrum (dash-dotted) at late times in run $C_{x1}$. 
    (b) Same for run $C_{x32}$.}
\label{fig:fig10}
\end{figure}

A similar behaviour is observed in the rotating case, as can be seen 
from Figs.~\ref{fig:fig8} and \ref{fig:fig9}. In these figures, the 
turbulent transport coefficients are computed from the runs in set 
$C$ ($\Omega = 4$, $R_o = 0.02$). As in the isotropic case, for 
sufficiently large $P_e$ turbulent diffusion remains approximately 
constant and the effective Schmidt number is of order one. $P_e$ 
number effects are observed earlier for vertical diffusion at small 
$P_e$ (indeed, for $P_e \le 500$ a change in the behaviour of 
${\cal D}_z$ can already be seen), although it seems that much 
smaller $P_e$ numbers are required to see a significant decrease in 
the horizontal diffusion. This could be associated with the fact that 
in the presence of rotation the turbulent diffusivity is already
reduced, and that therefore larger molecular diffusivities are needed
to decrease it further. Also, note that in the runs with large enough 
$P_e$ and small enough $R_o$, the vertical turbulent diffusion 
reaches a value that is approximately twice the horizontal, in good 
agreement with predictions in \cite{Cambon04, Vassilicos2}.

\begin{figure}
\centering
\includegraphics[width=4.5cm]{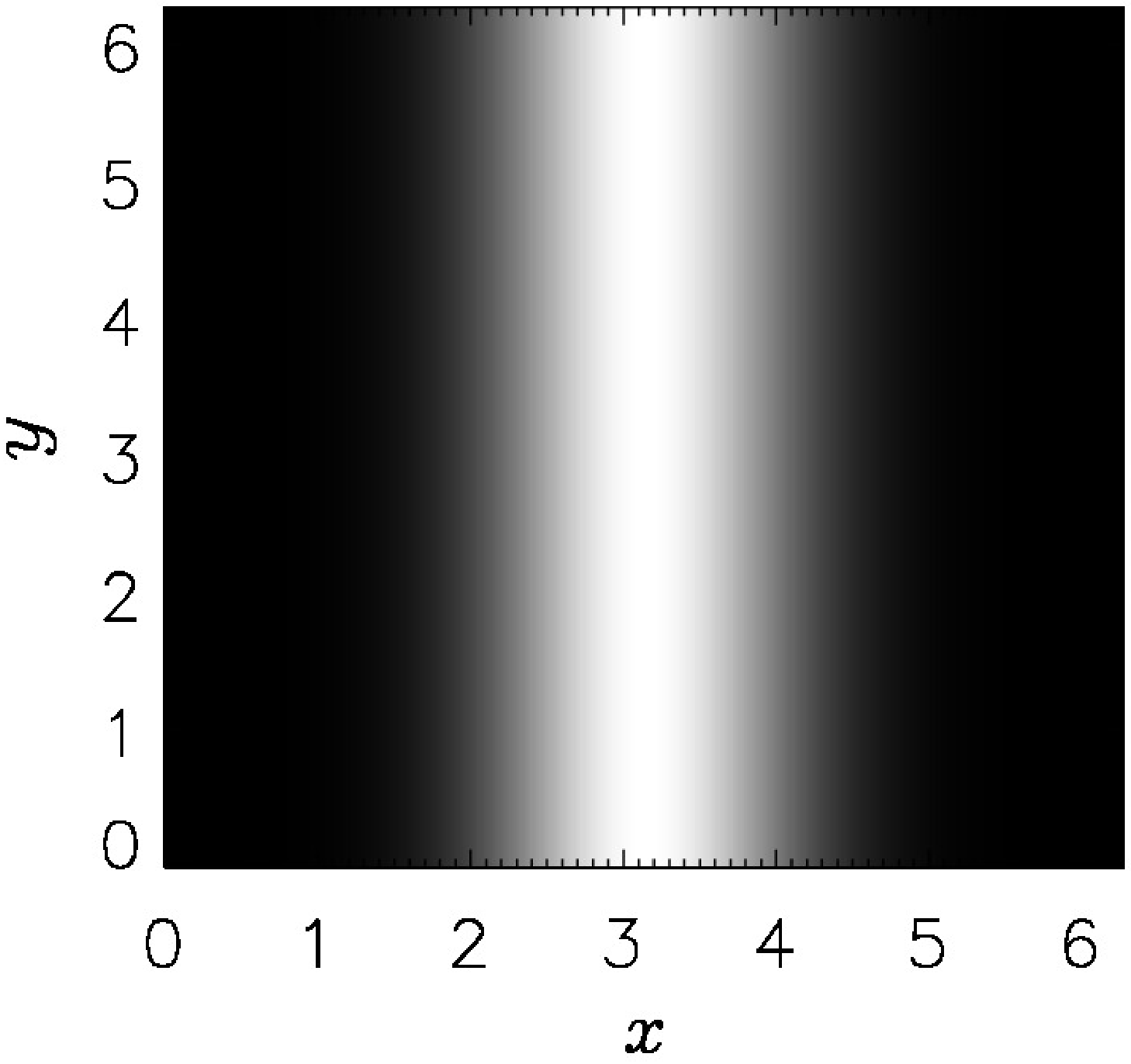}\includegraphics[width=4.5cm]{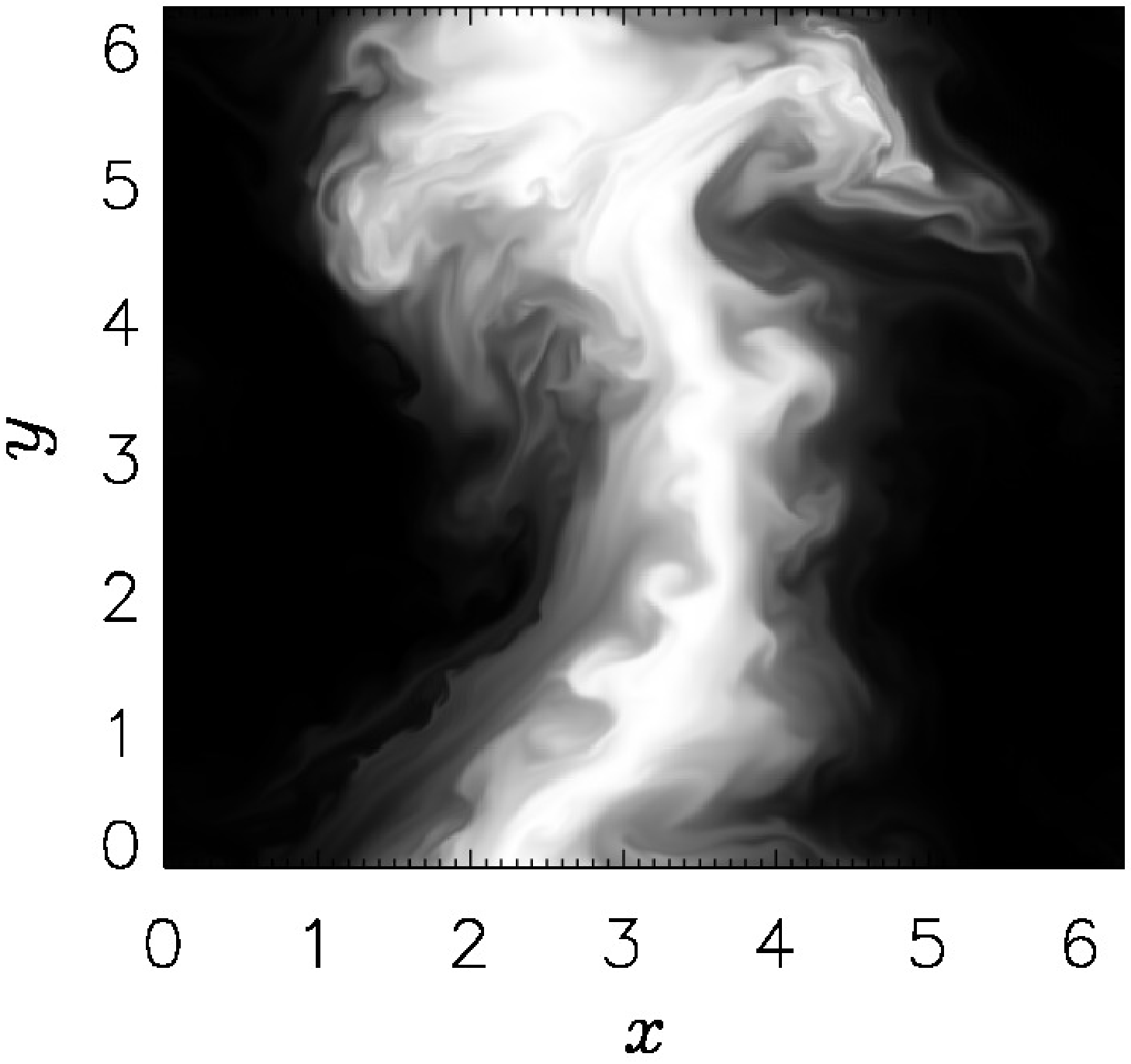} \\
\includegraphics[width=4.5cm]{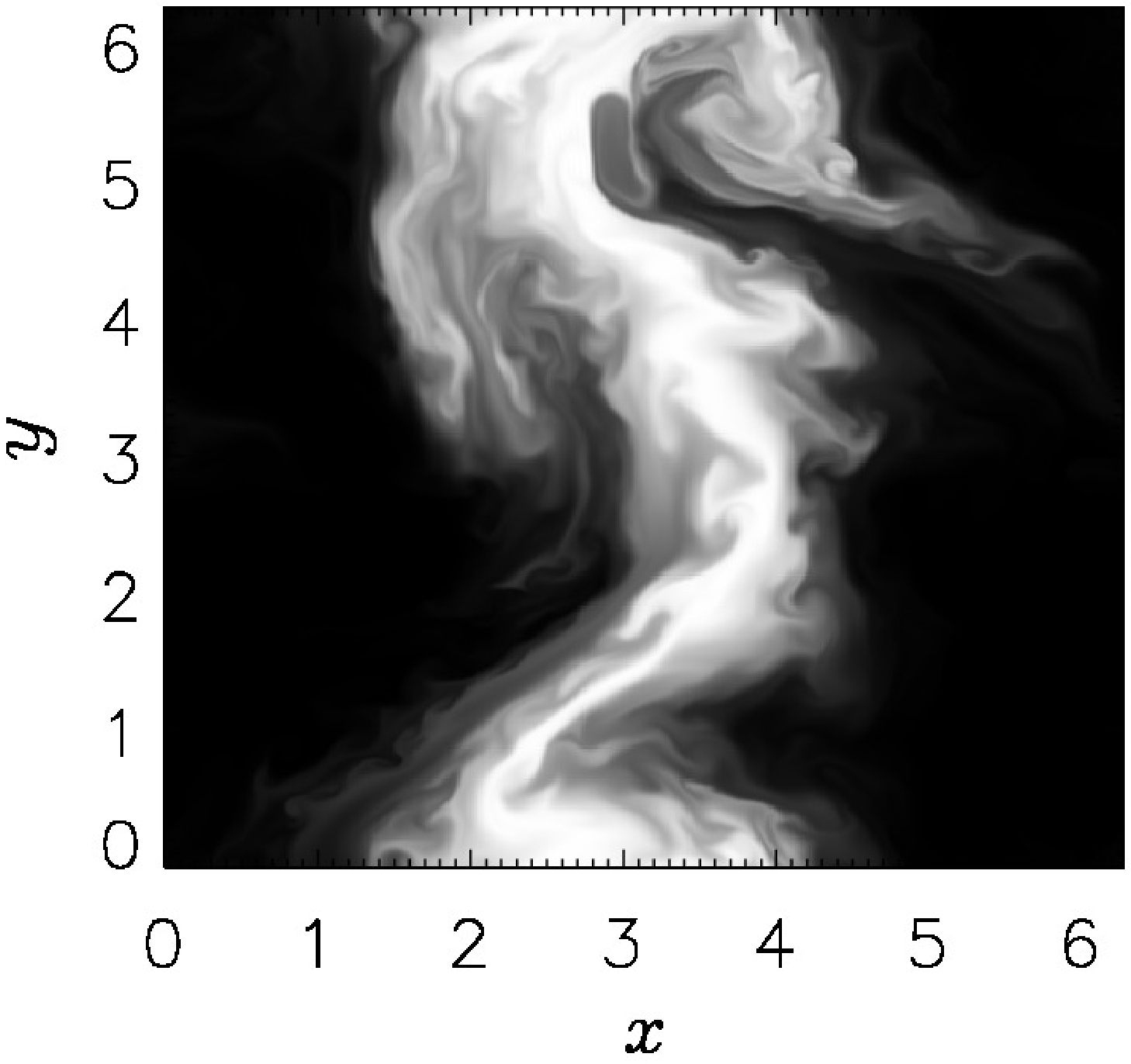}\includegraphics[width=4.5cm]{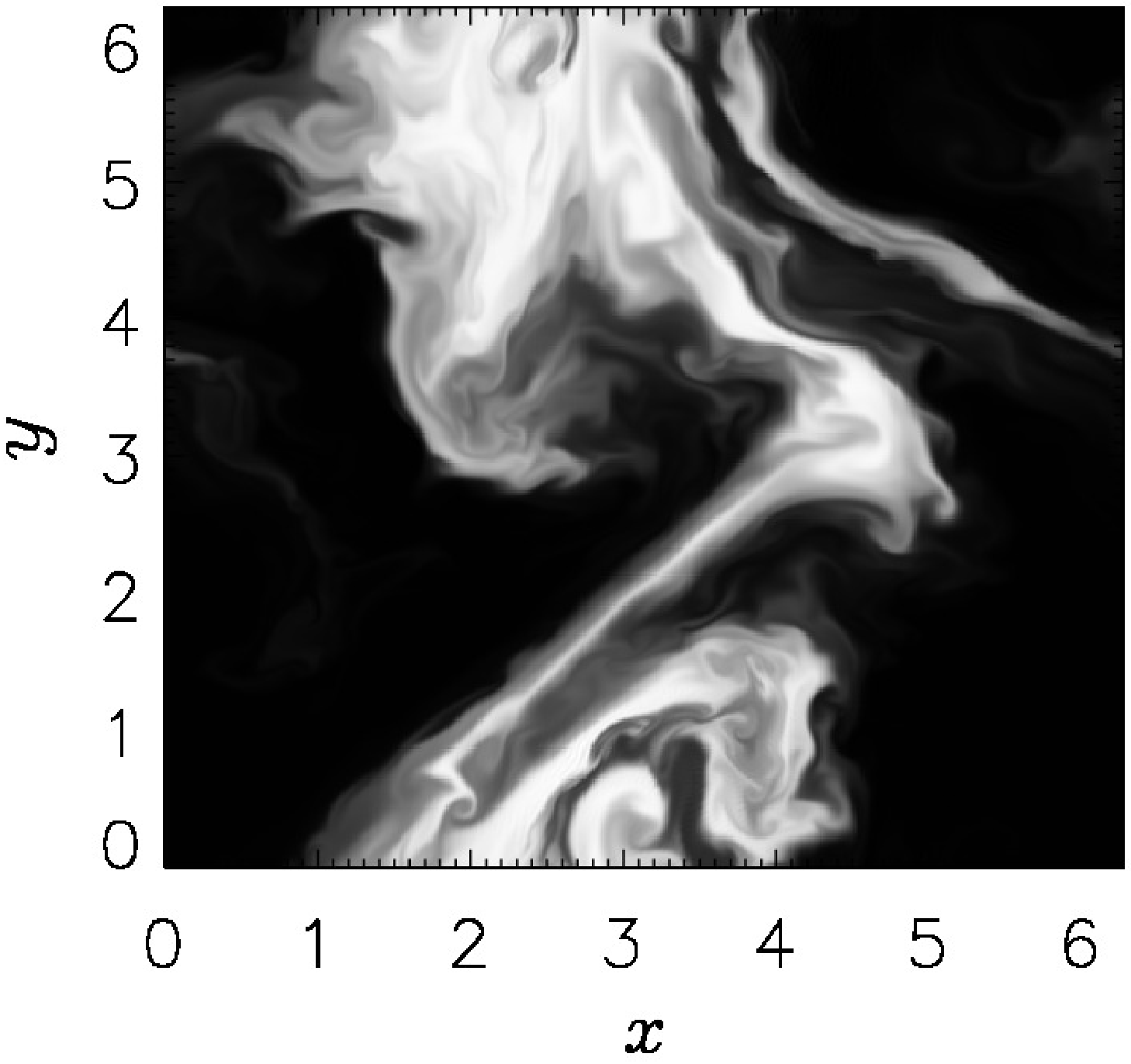}
\caption{Passive scalar concentration in a horizontal plane in run
  $C_{x1}$, at times $t = 0, 1, 1.5,$ and $2.5$ from left to right and
  from top to bottom.}
\label{fig:fig11}
\end{figure}

The change in the turbulent diffusivity as the $P_e$ number is
decreased can be also understood from the spectra of energy and 
passive scalar variance (see Fig.~\ref{fig:fig10}). For $S_c=1$, at
late times the energy and the passive scalar variance display an 
inertial range in a similar range of wave numbers. As a result, it 
can be expected that the turbulent eddies will mix the passive 
scalar concentration at all scales. For small $S_c$ and $P_e$, the 
passive scalar cannot develop an inertial range. As a result, eddies 
smaller than the dissipation scale for the passive scalar (i.e.,
eddies in the inertial range of the velocity field) are unable to mix 
the passive scalar and thus the turbulent diffusion decreases.

\subsection{Turbulent structures and diffusion}

\begin{figure}
\centering
\includegraphics[width=4.5cm]{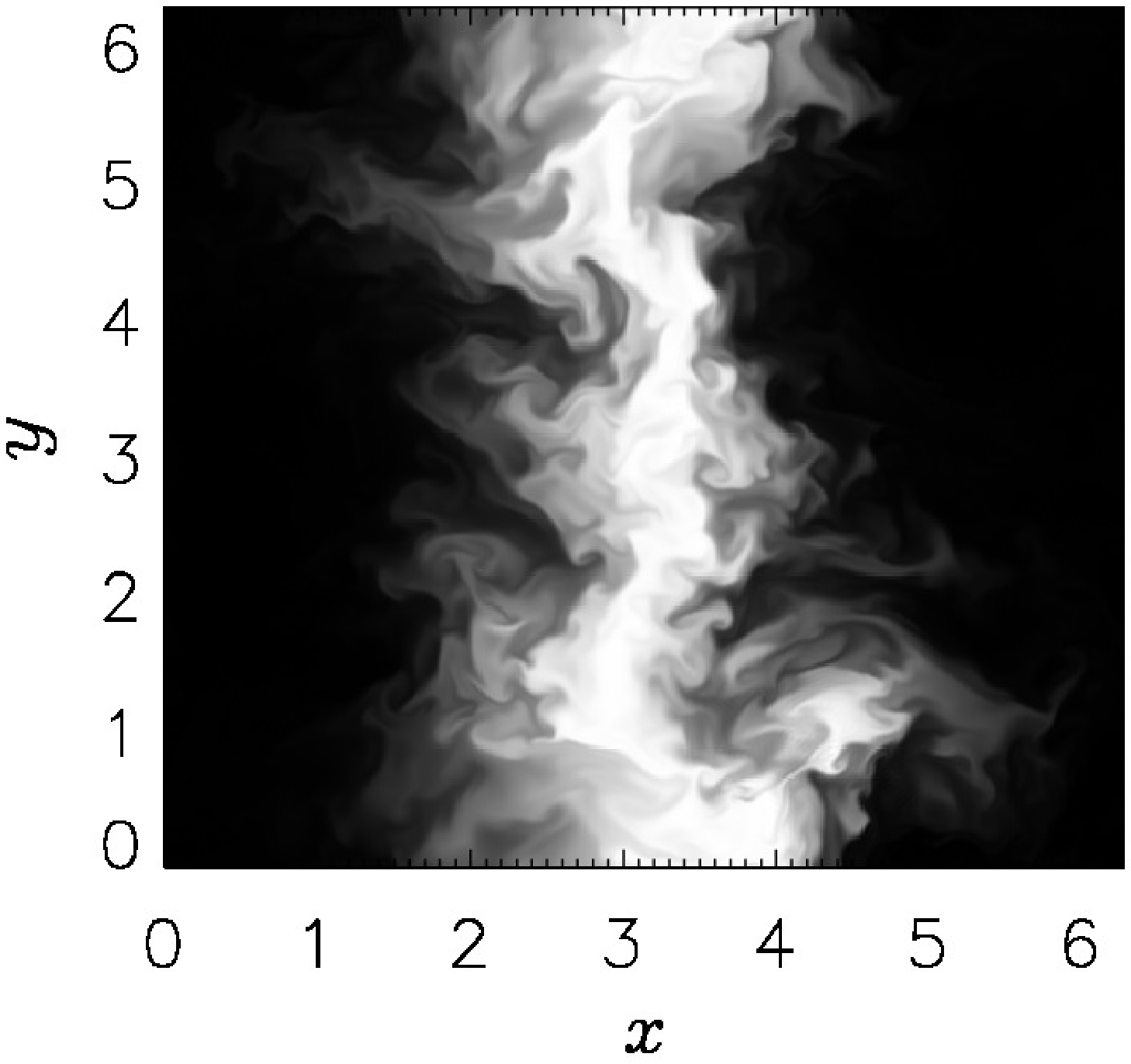}\includegraphics[width=4.5cm]{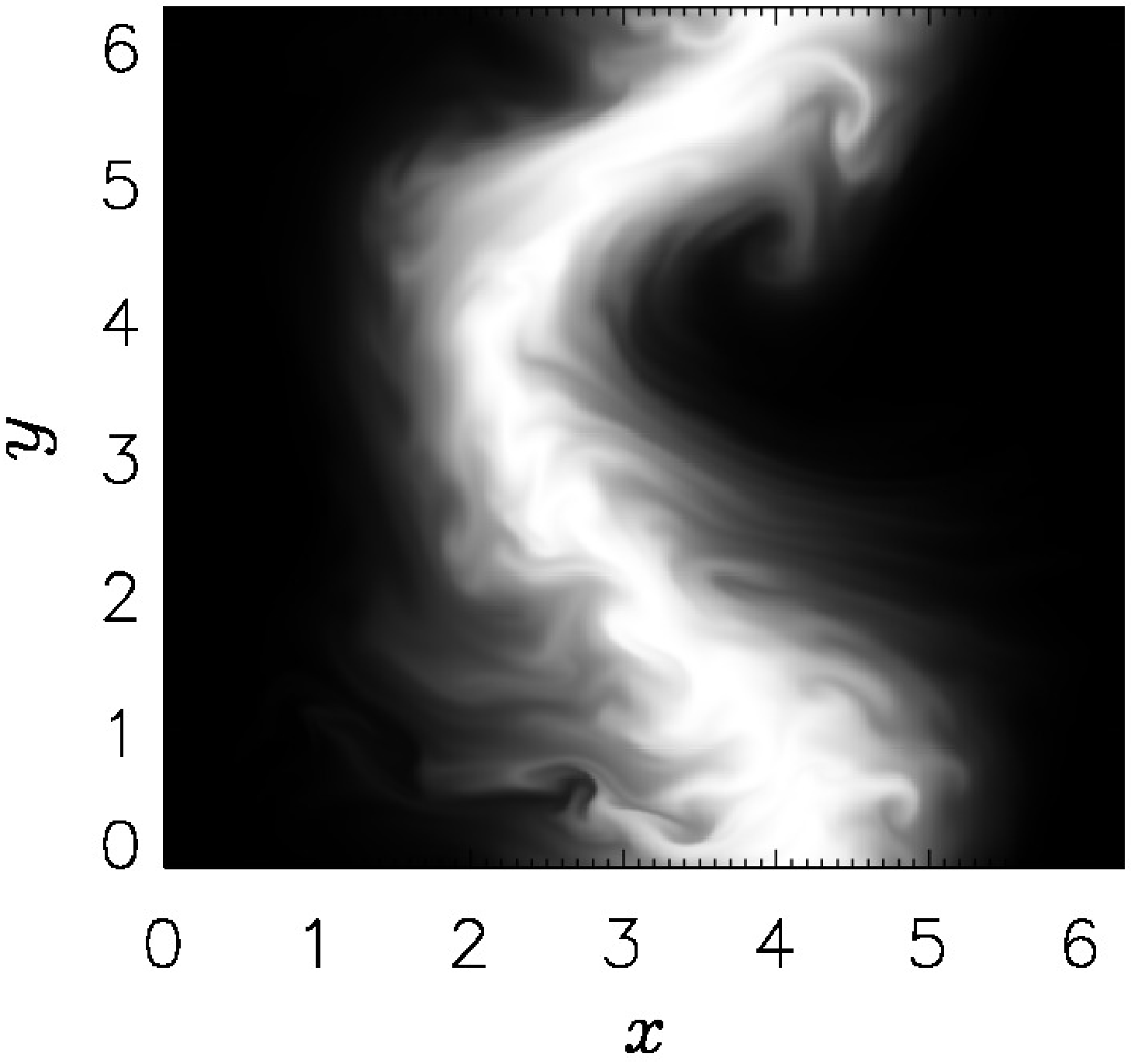}
\caption{Passive scalar concentration in a horizontal plane at $t=1$ 
 in runs $A_{x1}$ (left, no rotation) and $E_{x1}$ (right,
 $\Omega=8$).}
\label{fig:fig12}
\end{figure}

From the results presented above it is clear that rotation plays an 
important role in the diffusion of the passive scalar, modifying its
mixing with respect to the isotropic and homogeneous case. Figure 
\ref{fig:fig11} shows a cut in a horizontal plane of the passive
scalar concentration in run $C_{x1}$ at different times. The initial
Gaussian profile not only diffuses in time, but also bends and
rotates. The bending of the passive scalar concentration was observed 
before in \cite{Branden} and explained as an effect of the Coriolis
force. In our runs, the passive scalar at $t=0$ is concentrated in a 
narrow band around $x=\pi$. The average flux is thus towards positive
values of $x$ for $x>\pi$, and towards negative values of $x$ for 
$x<\pi$ (i.e., in the direction of $-\nabla \theta$, see 
Fig.~\ref{fig:fig2}). The Coriolis force in Eq.~(\ref{eq:NS}) is 
$-2 \Omega \hat{z} \times {\bf u}$ and therefore on the average this 
force creates a drift of the flux towards positive values of $y$ in
the $x>\pi$ region, and towards negative values of $y$ for $x<\pi$ 
\cite{Branden}. This explains the bending of the initial profile we 
observe of the runs with rotation, that is not observed in the runs 
without rotation (see Fig.~\ref{fig:fig12} for a comparison).

Diffusion in the parallel direction is of a different nature, and more 
strongly dependent of the structures that emerge in rotating turbulent 
flows. Rapidly rotating flows are characterized by columnar structures 
in the velocity field and vorticity, associated with a 
quasi-bidimensionalization of the flow. The mechanism underlying the 
transfer of energy towards two dimensional modes and responsible for 
the formation of these columns seems to be associated with wave 
resonances in the energy-exchanging triadic interactions
\cite{Waleffe}. Two-point closures of turbulence, such as the Eddy 
Damped Quasi-Normal Markovian closure (see, e.g., \cite{Cambon}) 
successfully explain the emergence of columns with the same
principle. However, there are alternative theories that consider the 
formation of columns as the result of a relative concentration of 
kinetic energy in cylindrical structures resulting from the
conservation of linear and angular momentum \cite{Davidson}.

\begin{figure}
\centering
\includegraphics[width=4.5cm]{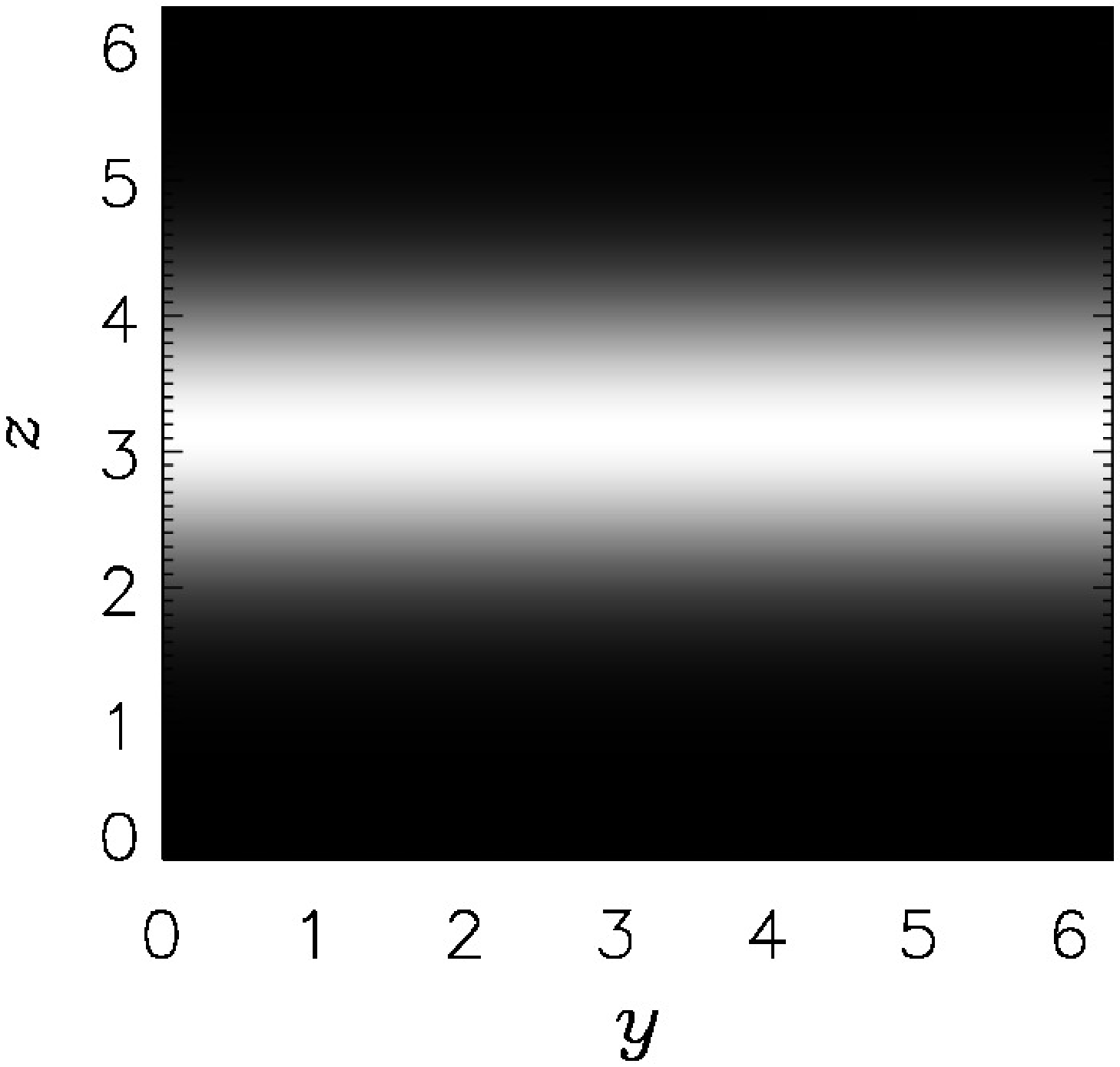}\includegraphics[width=4.5cm]{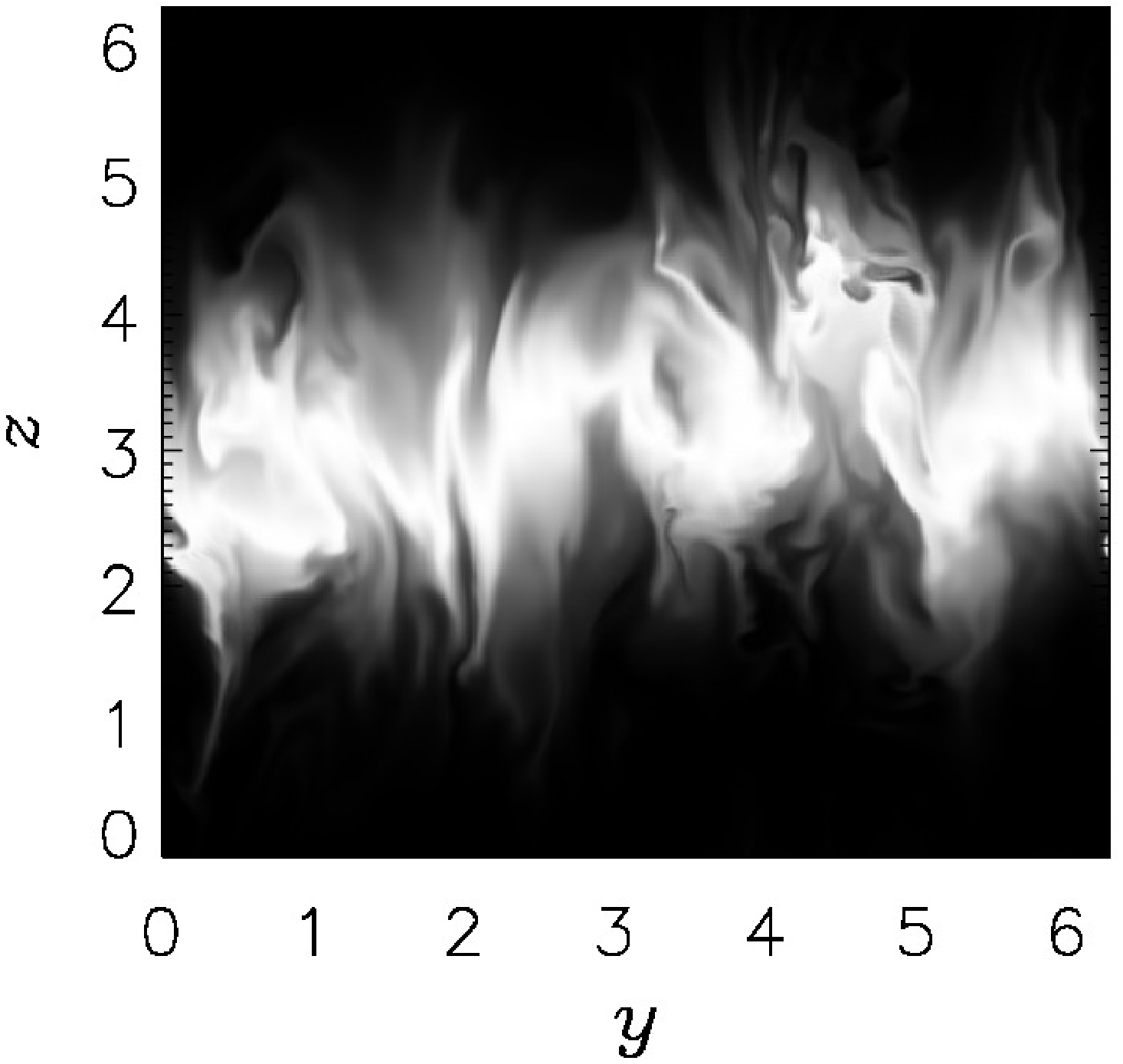} \\
\includegraphics[width=4.5cm]{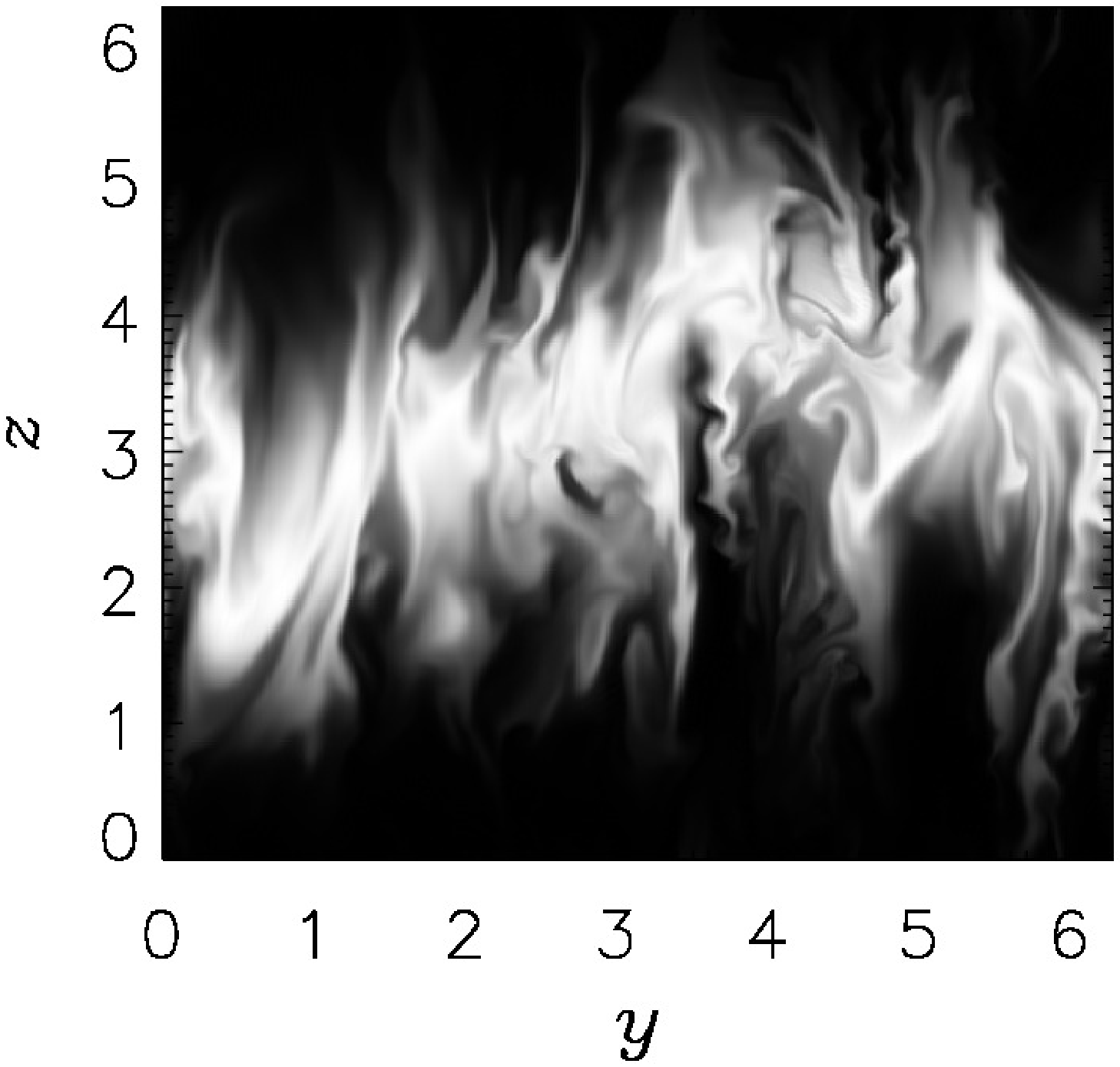}\includegraphics[width=4.5cm]{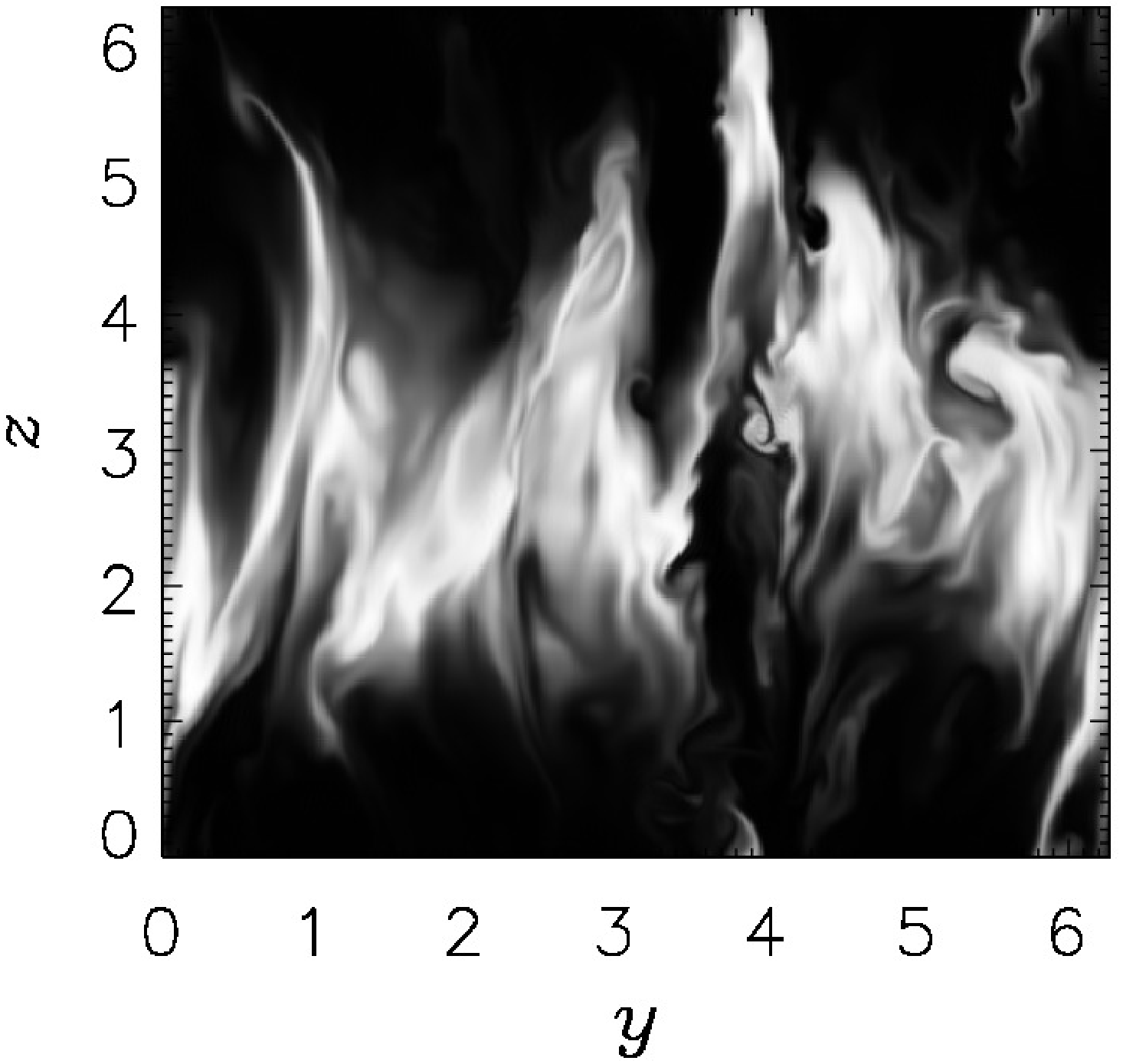}
\caption{Passive scalar concentration in a vertical plane in run
  $C_{z1}$, at times $t = 0, 1, 1.5,$ and $2.5$ from left to right and
  from top to bottom.}
\label{fig:fig13}
\end{figure}

Columnar structures have been reported in many numerical simulations 
of turbulent flows (see, e.g., \cite{Alexakis}). As these columns live for
long times and move across the domain, they play an important role in 
the mixing of the passive scalar. Figure \ref{fig:fig13} shows a cut
in a vertical plane of the passive scalar concentration at different
times in run $C_{z1}$. Note that diffusion is different from the one
observed in horizontal planes in the same run (Fig.~\ref{fig:fig11}),
and from the one observed in the isotropic and homogeneous case 
(Fig.~\ref{fig:fig12} (a)). The passive scalar is diffused from its
initial profile in vertical stripes, that are stretched further (thus
increasing the mixing) as time evolves. This stripes are created by
updrafts or downdrafts inside the columns. As these columns go 
through the region with large concentration of the passive scalar, the 
updrafts or downdrafts mix the passive scalar with the regions 
immediately above or below.

\section{Conclusions}

We used 56 direct numerical simulations with regular spatial 
resolution of $512^3$ grid points to measure turbulent diffusion in 
directions parallel and perpendicular to the rotation axis, in
turbulent flows at different Rossby and Schmidt numbers. 
The effective coefficients were obtained by studying the diffusion of 
an initial concentration of the passive scalar and calculated by 
measuring its average concentration and average spatial flux.

The effect of rotation in turbulent diffusion is the opposite of that
found in the presence of stratification \cite{Meneguzzi,Kimura1}:
while in the latter case stratification reduces vertical diffusion
with respect to horizontal diffusion, rotation dramatically reduces 
scalar diffusivity in the horizontal direction. In our simulations,
vertical diffusion remains of the same order as in isotropic and
homogeneous turbulence, although for a different reason: in rotating
flows, it was found that diffusion in the vertical direction is strongly 
dominated by updrafts and downdrafts in columnar structures
in the velocity field.

Within error bars, and for small enough Rossby number and large
enough Schmidt number, our results are consistent with theoretical 
results based on single-particle dispersion that predict that vertical 
diffusion is twice larger than horizontal diffusion in the presence of 
pure rotation \cite{Cambon04,Vassilicos2}. For small Schmidt and 
P\`eclet numbers the turbulent diffusion decreases as molecular 
diffusion becomes more important, while for large enough P\`eclet 
numbers the effective diffusion becomes independent of the P\`eclet 
number, resulting in a turbulent Schmidt number of order one.

\begin{acknowledgments}
The authors acknowledge support from grants No.~PIP 
11220090100825, UBACYT 20020110200359, and PICT 2011-1529 and 
2011-1626. PDM acknowledges support from the Carrera del 
Investigador Cient\'{\i}fico of CONICET. 
\end{acknowledgments}

\bibliography{ms}

\end{document}